\documentclass[conference]{IEEEtran}
\IEEEoverridecommandlockouts
\usepackage{comment}
\usepackage{cite}
\usepackage{amsmath,amssymb,amsfonts}
\usepackage[ruled,vlined,linesnumbered]{algorithm2e}
\usepackage{graphicx}
\usepackage{textcomp}
\usepackage{xcolor}
\usepackage{amsmath}
\usepackage{caption}
\usepackage{subcaption}
\usepackage[normalem]{ulem}
\usepackage{soul}
\usepackage{booktabs}

\def\BibTeX{{\rm B\kern-.05em{\sc i\kern-.025em b}\kern-.08em
    T\kern-.1667em\lower.7ex\hbox{E}\kern-.125emX}}
    
\usepackage{fancyhdr}
\usepackage{kantlipsum}

\newcommand{\fixme}[1]{{\color{red}\em\bf{[FIXME: #1]}}}
\fancypagestyle{plain}{
\fancyhf{}
\fancyhead[C]{Conference on \LaTeX} 

}
\usepackage{eso-pic}
\begin{document}

\AddToShipoutPictureBG*{
\AtPageUpperLeft{

\setlength\unitlength{1in}

\hspace*{\dimexpr0.5\paperwidth\relax}

}}

\title{Do Facial Trait Correlates with Roll Call Voting in Parliament? Using fWHR to Study Performance in Politics}

\author{\IEEEauthorblockN{Rahul Goel}
\IEEEauthorblockA{
\textit{Institute of Computer Science}\\
University of Tartu, Estonia \\
rahul.goel@ut.ee}
\and
\IEEEauthorblockN{Tymofii Brik}
\IEEEauthorblockA{\textit{Kyiv School of Economics} \\
Kyiv, Ukraine\\
tbrik@kse.org.ua}
\and
\IEEEauthorblockN{Rajesh Sharma}
\IEEEauthorblockA{\textit{Institute of Computer Science} \\
University of Tartu, Estonia\\
rajesh.sharma@ut.ee}

}

\maketitle
\IEEEoverridecommandlockouts
\IEEEpubidadjcol
\begin{abstract}  
Research has shown that people recognize and select leaders based on their facial appearance. However, considering the correlation between the performance of leaders and their facial traits, empirical findings are mixed. This paper adds to the debate by focusing on two previously understudied aspects of facial traits among political leaders: (i) previous studies have focused on electoral success and achievement drive of politicians omitting their actual daily performance after elections; (ii) previous research has analyzed individual politicians omitting the context of social circumstances which potentially influence their performance. We address these issues by analyzing Ukrainian members of parliament (MPs) who voted for bills in six consecutive Verkhovna Rada starting from Rada 4 (2002-06) to Rada 9 (2019-Present) to study politicians' performance, which is defined as co-voting or cooperation between MPs in voting on the same bill. In simple words, we analyze whether politicians tend to follow leaders when voting. This ability to summon the votes of others is interpreted as better performance. To measure performance, we proposed a generic methodology named Feature Importance For Measuring Performance (FIMP) that can be used in various scenarios. Using FIMP, our data suggest that MPs vote has no impact from their colleagues with higher or lower facial width-to-height ratio (fWHR), a popular measure of the facial trait. 

\end{abstract}


\section{Introduction}\label{sec:introduction}

Research has shown that individuals' facial traits are associated with men's testosterone levels \cite{lefevre2013telling} and predict aggressive \cite{carre2008your} and risk-taking behavior \cite{haselhuhn2012bad}. At the same time, psychological studies show that people judge others by their facial appearance \cite{todorov2015social,zhao2003face}. People infer whether a person is likely to be trustworthy, competent, or dominant by looking at photos or computer-generated images \cite{buckingham2006visual}. However, research also points out that the process of social attribution from faces is very complicated. In real life, people rely on multiple sources of information about a person \cite{todorov2015social}, while also being affected by biases, stereotypes, and information overload \cite{stoker2016facial}. This complexity brings many challenges to empirical studies of leadership. On the one hand, there is evidence that facial traits are associated with better performance in arts \cite{lebuda2016written} and athletics \cite{welker2015examination}. On the other hand, other studies show that facial appearance does not matter for CEOs' performance in business \cite{stoker2016facial}. Therefore, there is still a demand for empirical evidence whether facial traits typically associated with leadership in a competitive environment (e.g., dominance or masculinity) are relevant for actual leader ability and performance.

This paper adds to the debate by analyzing the performance of 
members of the Ukrainian parliament (MPs). Politicians' performance is interpreted as a collaboration between MPs when they cast the same vote on the same bill. Such collaboration is critical since the success of a bill is a result of collective action. While co-voting can be treated as a signal about coalition building \cite{aleman2009comparing}, failure to collaborate can signal distortion and weak parliament \cite{magelinski2019detecting}. Recent studies in political science acknowledge that voting behavior is influenced by socializing in the parliament and informal ties \cite{canen2019endogenous,mcclurg2014political,zhang2008community}. Therefore, influencing fellow politicians to cast a vote for the same bill is crucial for a political leader. Moreover, in the context of the Ukrainian parliament, where politicians often change their stance on ideological spectrum and recruit votes from different parties \cite{magelinski2018legislative,magelinski2019detecting, harder2019hybrid, harder2020predicting}, this ability to influence co-voting becomes particularly important. Thus, we treat co-voting as a measure of performance. 


In this work, to study the correlation between MPs' facial traits and their performance, we downloaded photos of all the MPs from the Ukrainian Parliament's official website to measure their masculinity using the facial width-to-height ratio (fWHR). 
Please note that the photos collected are publicly available on the official website of the Parliament of Ukraine, and their use for research purposes is not in violation of a privacy issue. For our work, the use of the fWHR index is reasonable as broad face looking people tend to show more physical strength \cite{sell2009human,marcinkowska2019women,windhager2011geometric,fink2007male}, 
competitive ability \cite{dixson2014eye}, behavioral dominance \cite{geniole2015evidence}, and higher social rank in hierarchies \cite{hill2013quantifying}. 
The fWHR is then compared against their similarities in roll-call voting for bills from Rada 4 to 9.

\textbf{Research question and hypotheses: }Building on the literature in social networks \cite{jackson2010social,canen2019endogenous, zhang2008community} and previous studies of fWHR \cite{stoker2016facial} we expect that members of parliament (MPs) are likely to be influenced by their colleagues with leadership qualities. Our research question is \textbf{to what extent fWHR influence roll call voting of politicians}. Below we consider several hypotheses to guide our empirical analysis:

\begin{enumerate}
    \item Social network theory suggests that people tend to create ties with those similar to them, which is known as the homophily effect \cite{mcpherson2001birds}. Thus one could expect clusterization of MPs by fWHR (lower with lower and higher with higher).
    \item On the other hand, literature in social psychology suggests that people tend to follow leaders \cite{van2015many}. Thus, one could expect a preference of MPs with lower fWHR to follow MPs with higher fWHR when voting.
    \item Finally, the literature in political science suggests that party affiliation and ideological differences are far more important than other variables \cite{poole1985spatial, andris2015rise, neal2020sign}. This role of affiliation has also been registered among respondents who tend to favor leaders that look like members of their preferred political party \cite{olivola2012republicans}. Moreover, people judge others based on stereotypes (age, gender, race, social status) \cite{king2006s}. Therefore, the role of fWHR in the parliament could disappear after controlling for other social variables. 
\end{enumerate}

In what follows, we first review relevant literature to elaborate on these hypotheses and then discuss our data and methods to address them. Our analysis suggests that fWHR has no role in driving legislative co-voting. Ukrainian MPs were likely to support legislative bills supported by their colleagues irrespective of their fWHR index. The findings remained the same after controlling for different types of legislative bills and simulating the data. 

The contribution of our analysis is four-fold.
\begin{enumerate}
    \item Previous studies of political leadership focused on either electoral success or personal traits of politicians \cite{berggren2010looks,todorov2005inferences,lewis2012facial}. In contrast, this paper looks at politicians' performance in the real-life scenario of roll-call voting.
    \item Second, while previous studies relied on respondents \cite{berggren2010looks,todorov2005inferences}, we use a data-driven approach to measure the facial trait of politicians.
    \item  Third, we bridge the psychological literature of leadership and facial trait \cite{todorov2015social,zhao2003face} with the sociological literature of social networks \cite{jackson2010social}.
    \item Finally, we proposed a generic methodology named Feature Importance For Measuring Performance (FIMP) that can be used to test feature as performance measure.
\end{enumerate}
\section{Related Work}\label{sec:relatedWork}

\subsection{Theoretical background and empirical findings}

Research of human skulls suggests that males have a larger relative facial width than females. This sexual dimorphism is interpreted as an honest signal of masculinity, aggression, and related traits \cite{kramer2012lack}. Simultaneously, the same research cast doubts about whether faces (in contrast to skulls) display the same difference. Despite existing criticism \cite{hodges2016facial} and challenges to study faces of women \cite{lefevre2013telling}, the relative facial width is considered to be a good predictor of masculine behavior associated with strength, aggression, and dominance \cite{carre2008your, haselhuhn2012bad, lewis2012facial, wong2011face}. At the same time, humans judge others by looking at their faces or photos. Research has shown that dominant-looking individuals are more likely to be judged as leaders by respondents. Voters tend to favor leaders with masculine faces if selecting a wartime leader \cite{little2007preferences, spisak2012warriors}. Moreover, respondents can infer the leadership domain from facial appearance \cite{olivola2014many}. The evolutionary perspective suggests that particular facial traits were associated with leadership in ancestral human environments,  and this tendency to judge leadership from faces is still present in modern societies \cite{van2015many}. How people judge faces has practical consequences for how societies are organized since members of respective communities often select leaders. Research has shown that candidates' dominant facial traits influence the selection of leaders, e.g., promotion in business \cite{alrajih2014increased, stoker2016facial} and electoral support in politics \cite{berggren2010looks, olivola2010elected}. Moreover, facial traits predicts people's performance, such as success in sports \cite{welker2015examination} and even in arts \cite{lebuda2016written}.

However, little is known about the actual performance of political leaders after elections. To what extent facial trait influence their daily work in the parliament? Does it impact the voting process? Political science has long addressed polarization of legislators \cite{poole1985spatial, andris2015rise, neal2020sign}. Recently, a new wave of scholarship emerged in political science with a strong focus on social network analysis, showing that politicians' voting is influenced by their ties to other politicians and their embeddedness in more extensive social networks \cite{canen2019endogenous, zhang2008community}. Thus, how politicians interact with other politicians when making their voting decisions is crucial. There are some studies on nonverbal communication in politics \cite{dumitrescu2016nonverbal} and the role of facial hair in roll‐call voting \cite{herrick2015razor}. However, the link between fWHR and co-voting is understudied. 

\subsection{Data-driven approach to measure facial trait in politics}


Relative facial width can be inferred from people's photos using the facial width-to-height ratio (fWHR) \cite{stoker2016facial}. Width is typically measured by the distance between the left and the right zygion (bizygomatic width). At the same time, height is the distance between the upper lip and the midpoint of the eyebrows' inner ends (upper facial height) \cite{weston2007biometric}. This metric has been widely applied in social psychology, and other fields of social science \cite{stoker2016facial}. For example, research of 29 former US presidents has shown that their fWHR is positively associated with their achievement drive \cite{lewis2012facial}.

\subsection{Summary and justification of our analysis}

Previous studies of leadership in various domains showed that facial trait predicts performance in athletics and arts \cite{welker2015examination,lebuda2016written}. Considering other fields, facial trait predicts CEOs' selection in business, however, it does not predict their performance \cite{stoker2016facial}. When people are embedded in the context of particular industries, social circumstances, and institutions, their individual traits are downplayed. Little is known whether facial trait predicts the actual performance of politicians in the context of co-voting. 
When politicians support the same bill, it is more likely that this bill will pass. Therefore, political leaders should be able to influence other politicians to vote with them. On the other hand, ideological differences, strategic interests, and socializing in parliament could be more critical than leader ability \cite{canen2019endogenous, neal2020sign}. Moreover, a general tendency of social networks to create clusters of similar agents (also known as homophily \cite{mcpherson2001birds}) could lead to politicians' clusterization by fWHR: low with low and high with high. This paper looks into empirical data of roll-call voting for several years among Ukrainian politicians to provide new evidence of whether facial trait predicts politicians' performance.

\section{Dataset Description}\label{sec:datasetDescription}
This section describes the voting mechanism of the Ukrainian parliament and the datasets used in this paper.

\begin{table*}
\centering
\begin{tabular}{lllllll}
\toprule
                            & Rada 4            & Rada 5             & Rada 6             & Rada 7             & Rada 8             & Rada 9              \\
                            \midrule
Time Period                 & 2002-06           & 2006-07            & 2007-12            & 2012-14            & 2014-19            & 2019-Present        \\
\midrule
   \multicolumn{7}{@{}l}{\textbf{MP Characteristics}}\\
Number of MPs               & 505               & 478                & 539                & 479                & 469                & 435                 \\
Sessions                    & 9                 & 4                  & 10                 & 5                  & 10                 & 3                   \\
Number of Parties           & 10                & 4                  & 4                  & 8                  & 17                 & 15                  \\
Number of Fraction          & 20                & 6                  & 9                  & 9                  & 9                  & 8                   \\
MPs mean Age                & 48.73             & 49.27              & 48.62              & 48.14              & 44.39              & 44.24                    \\
MPs sd Age                  & 10.25             & 10.35              & 10.43              & 10.45              & 9.72               & 9.65                    \\
Male/Female ratio           & 17.70        & 10.66        & 11.81        & 9.39       & 7.25               & 3.71        \\
\midrule
   \multicolumn{7}{@{}l}{\textbf{Rollcall Characteristics}}\\
Number of Bills             & 10929             & 3424               & 23282              & 4135               & 24103              & 3857                \\
Bill Passed Ratio           & 0.38       & 0.64        & 0.21          & 0.32        & 0.12        & 0.35         \\
Yes \%                      & 43.68             & 50.37              & 34.64              & 45.86              & 28.09              & 33.26               \\
No \%                       & 0.87              & 0.4                & 0.35               & 2                  & 0.9                & 6.05                \\
Did not vote \%             & 36.06             & 26.54              & 36.7               & 22.91              & 33.48              & 14.5                \\
Abstain \%                  & 0.43              & 0.15               & 0.1                & 1.74               & 5.7                & 31.26               \\
Absent \%                   & 18.96             & 22.54              & 28.2               & 28.51              & 31.83              & 14.48 \\
\midrule
   \multicolumn{7}{@{}l}{\textbf{Facial Traits}}\\
MPs mean fWHR               & 1.92          & 2.03           & 1.96           & 1.96           & 1.99           & 2.01           \\
MPs sd fWHR                 & 0.14       & 0.15        & 0.14        & 0.13        & 0.14          & 0.15        \\
D'Agostino's K-squared test & No                & Yes                & Yes                & Yes                & Yes                & Yes                 \\
Jarque–Bera test            & No                & Yes                & Yes                & Yes                & Yes                & Yes                 \\
Kolmogorov–Smirnov test     & Yes               & Yes                & Yes                & Yes                & Yes                & Yes                 \\
Shapiro-Wilk test           & No                & Yes                & Yes                & Yes                & Yes                & Yes                 \\
Conclusion                  & Weak & Strong & Strong & Strong & Strong & Strong  \\
\bottomrule
\end{tabular}
\caption{Dataset Description}
\label{table:DatasetAllRada}
\end{table*}
\subsection{The voting mechanism}

The Ukrainian Parliament or \textit{Verkhovna Rada}, is the legislative body that has received considerable attention in the scholarship due to the recent history of political turbulence in Ukraine \cite{harder2019hybrid, harder2020predicting, magelinski2018legislative, magelinski2019detecting}. Verkhovna Rada appoints the government, votes on national laws, and allocates the national budget. It was designed to include 450 MPs; however, the actual number has been 420 since the annexation of Crimea and the hybrid war with Russia in the Donbas region of Ukraine. For the last three elections, half of the seats have been filled by direct election and half by parties. Studies of recent convocations show that legislative polarization was visible among new MPs and independent MPs (i.e., those filled by direct elections), while party leaders maintained their own stances \cite{magelinski2019detecting, magelinski2018legislative}. Moreover, the same studies suggest that the bill was more likely to be supported if the President sponsored it. 

\subsection{The datasets}
This work utilizes four separate publicly available datasets; Rollcall, Bills identifier, Member of Parliament's (MP's) characteristics \cite{ULeNet} and MP's images which are available on the official website\footnote{iportal.rada.gov.ua}. For the rest of this work, we refer to them as \textit{Rollcall}, \textit{Identifiers}, \textit{MP's characteristics} and \textit{Facial trait}, respectively. Please note that in this study, we analysed six consecutive \textit{Verkhovna Rada} starting from Rada 4 (2002-06) to Rada 9 (2019-Present).

\noindent \textbf{\textit{1) Rollcall}} describes MP's voting behavior during \textit{Rada} for each bill. The data comprises votes information for MPs, with each row representing an MP and each column representing a bill. An MP can vote either ``Yes'', ``No'', ``Did not vote'', or ``Abstain'' for a bill. The absence of an MP is also indicated by the marking ``Absent'' in the data.

\noindent \textbf{\textit{2) Identifiers}} consist details of 77,673 bills from 12-May-1998 to 02-July-2020. This includes 11 different types of bills: \textit{Amendments} (24,073 bills), \textit{Final voting} (18,989), \textit{Not classified} (15,860), \textit{Agenda} (6,595), \textit{Short procedure} (5,888), \textit{Cancel} (2,756), \textit{Signal voting} (1,357), \textit{Second voting} (1,252), \textit{Revision} (602), \textit{President} (217) and \textit{Alternative} (84). In this work, we concentrate on Rada 4 (2002-06) to Rada 9 (2019-Present). The details about bills and distribution of votes during various Rada are shown in Table \ref{table:DatasetAllRada} under \textbf{Rollcall}. In Ukraine, when MPs do not support some legislation, they signal it either by not voting at all or abstaining. Therefore, this paper treats voting as a dichotomy of ``Yes'' and all other possible outcomes.

\noindent \textbf{\textit{3) MP characteristics}} contains each MP's features for the last seven convocations of the Ukrainian Parliament, including the present one (1998-present). Depending on the specific convocation, some additional information is also available for MPs. Overall, the dataset includes 3,368 observations and 31 features representing 2,040 unique MPs and two institutions - the President and Cabinet of Ministers. The features include the Ukrainian and English versions of MPs names, their gender, party, faction (which is the large group of MPs in parliament that might include different parties), and the way they were elected (party system or constituency). Depending on convocation, there is also additional information, for instance, education, previous occupation, or the number of assistants. Importantly, if the MP changed several factions during the convocation, only the last one is included in the dataset.
    
Following our problem, we extracted various features such as gender, education, date of birth, party, specialization, factions position, etc., for all MPs in various Rada. In Table \ref{table:DatasetAllRada} under \textbf{MP characterstics} number of MPs, sessions, number of fraction, parties, mean age, MPs male-female ratio are mentioned for each Rada. This can easily be observe that during Rada 4, male-female ratio was much higher compared to the present Rada, which indicate that female participation in Ukraine parliament is increasing with time.

\noindent \textbf{\textit{4) Facial trait}} is collected using web scrapping. Please note that the collected photos are publicly accessible on the official website of the Parliament of Ukraine, and the use of these images for research purposes is not against privacy concerns.

Previous studies suggested that one has to be very careful when working with images of faces \cite{stoker2016facial}. On manually reviewing MP's web-scrapped photos from the Parliament of Ukraine's official website, we observe that in some of the images, MP's are not looking directly into the camera or their faces are tilted. Therefore, we decided to do the quality assurance of each photo before performing further analysis. We begin by correcting the face alignment. 
Our goal is to warp and transform all images with same facial alignment using three necessary affine transformations:  rotation, translation, and scaling. 

\begin{figure}
    \centering
    \includegraphics[width=\columnwidth]{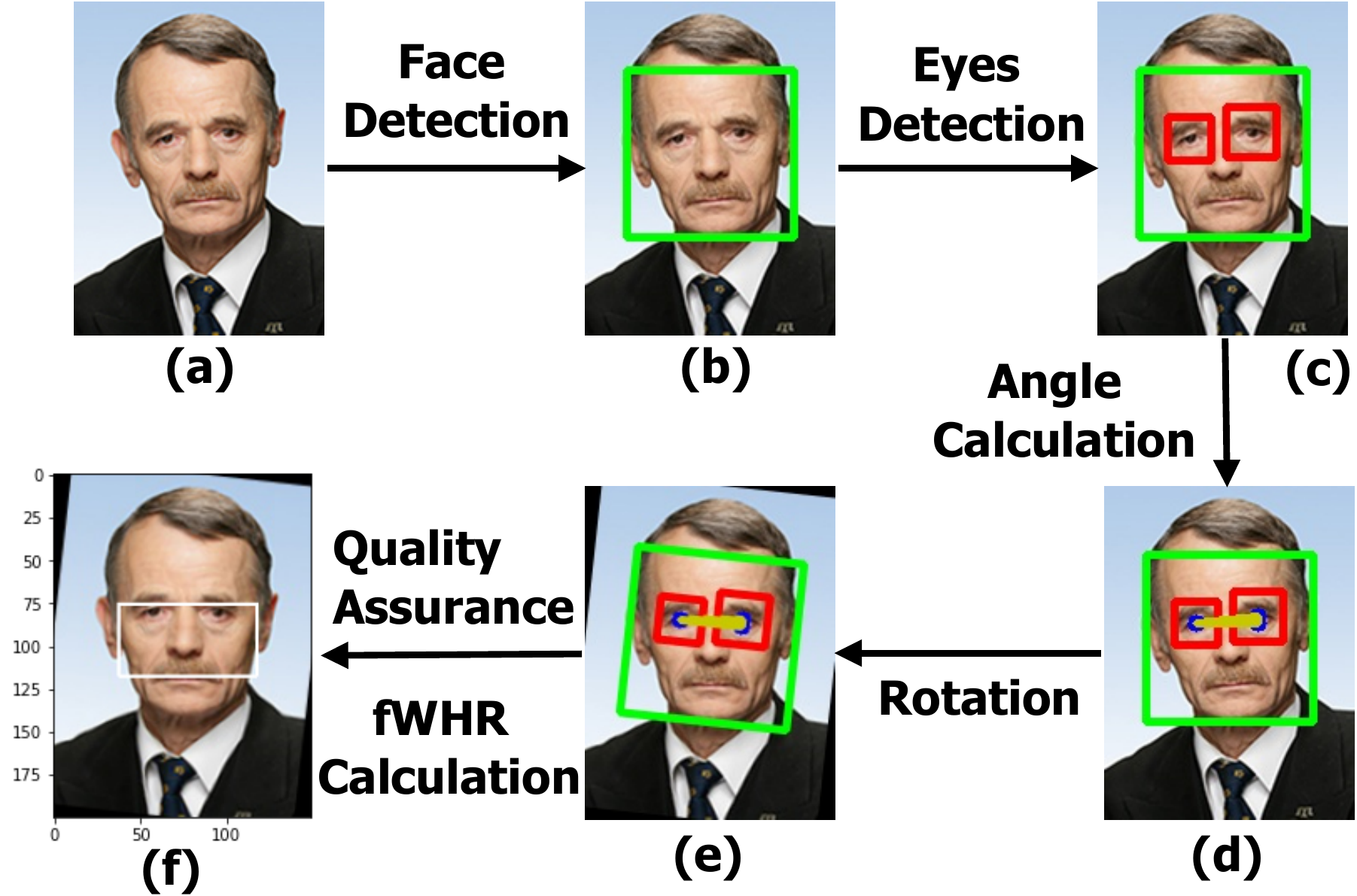}
    \caption{\textbf{Face Alignment Correction \& Image Quality Assurance: }In (a), we start by showing the web-scrapped photo of an MP from the Parliament of Ukraine’s official website. The face is highlighted using the green box in (b) once detected. Next, eyes are pointed out using the red box in (c). Based on the center point of the eyes, the angle of rotation is calculated (shown in (d)), and the image is rotated by the identified angle, as shown in (e). Finally, quality assurance is carried out to verify if the facial boundaries are clearly visible for the fWHR calculation (shown in (f)).}
    \label{fig:faceAlign}
\end{figure}

To perform face alignment, we used a method that focused on areas around the eyes. Please refer Figure \ref{fig:faceAlign} as an example for face alignment. This process consists of following steps: (1) \textbf{Detecting faces and eyes in the image}; (2) \textbf{Calculating the center of detected eyes}: Let, centre of left and right eyes are ($x_l$, $y_l$) and ($x_r$, $y_r$); (3) \textbf{Drawing a line between the center of two eyes}; (4) \textbf{Drawing the horizontal line between two eyes}; (5) \textbf{Calculating length of 3 edges of the triangle}; (6) \textbf{Calculating the angle}: The length of the base and height of the triangle are $\Delta x = x_r - x_l$ and $\Delta y = y_r - y_l$ respectively. So, the angle of rotation is $\Theta = arctan\frac{\Delta x}{\Delta y}$; (7) \textbf{Rotating image by calculated angle}; and (8) \textbf{Scale the rotated image}.

\begin{figure}
  \begin{tabular}{cc}
\subfloat[The fWHR value is 2.2.]{\includegraphics[width=0.45\columnwidth]{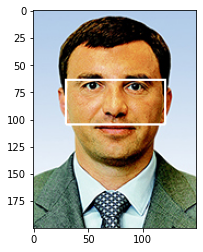}} &
\subfloat[The fWHR value is 1.86.]{\includegraphics[width=0.45\columnwidth]{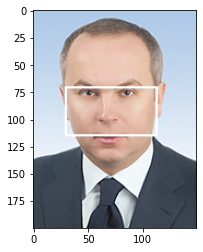}}
\end{tabular}
  \caption{Facial boundaries comparison.}\label{fig:MpsfWHR}
\end{figure}

Next, we calculated the facial width-to-height ratio (fWHR) using rotated photos of MPs in the \textit{Facial trait}. According to \cite{carre2008your}, the face's width is calculated as the maximum horizontal distance between the right and the left facial boundary; the height of the upper-face is measured as the vertical distance between the highest point of the upper lip and the highest point of the eyelids. The fWHR is then measured as the width divided by the height for MPs (with a clearly visible face) for the \textit{Rada}. Here, every image is again quality assured to see if it fits for fWHR calculation that is clear visibility of the facial boundary needed to calculate the fWHR. 
To exemplify, Figures \ref{fig:MpsfWHR}(a), and \ref{fig:MpsfWHR}(b) show two politicians with fWHR values of 2.2 and 1.86 respectively. In other words, the former MP is classified to have relatively higher fWHR since his facial boundaries are relatively larger than the height of the upper face. 

The details about quality images among all images, mean fWHR, fWHR standard deviation, various normality test, and their conclusion during various Rada are shown in Table \ref{table:DatasetAllRada} under \textbf{Facial Traits}. We can observe that mean fWHR over the years is similar. From normality test, we can conclude that Rada 5 to Rada 9 follows strong normal distribution. However, Rada 4 follows weak normal distribution.

Finally, after all data merging, we created a final data, which for convenience, we will label as \textit{MP-performance}. The Table \ref{table:DatasetAllRada} gives the dataset statistics and also highlights that the fWHR follows a normal distribution, which indicates that most of the Ukrainian MPs are moderate in terms of their facial masculinity, with some extreme examples on both sides of the scale. The fact the distribution of fWHR is normal also signals that the data are of relatively good quality, which does not underestimate or overestimate fWHR due to the bad quality of photos.

\noindent \textbf{Votes encoding scheme: }As we mentioned earlier, we analyze co-voting or the process when MPs cast support for the same bill. We follow previous studies \cite{magelinski2018legislative, magelinski2019detecting} and code all voting instances as binary events with ``Yes'' (supporting the bill) coded as 1 and everything else coded as 0. We are interested in MPs' ability to summon support to legislation among other politicians instead of any opposition. In the context of the Ukrainian parliament, MPs typically use one of the other voting options when they oppose the bill.
\section{Methodology}
This section explains the approach and challenges involved in identifying masculinity's effect on the voting behavior (Section \ref{subsec:method}) and our approach for overcoming these challenges (Section \ref{subsec:optimalK}).

\subsection{Identifying the impact of fWHR}\label{subsec:method}
In order to answer our key research question, that is, whether politicians follow high fWHR colleagues when voting, we adapt an approach that identifies the k-most similar MPs for each MP considering their voting behaviour on the \textit{MP-performance} dataset. The process is explained step-by-step in the Algorithm \ref{algo:HIM}, \textit{Feature Importance For Measuring Performance (FIMP)}. Given the \textit{MP-performance} matrix, \textit{V}; MPs fWHR value, \textit{M}$_p$; and \textit{K} as input in step 1, the \textit{FIMP} algorithm identifies for each value of \textit{i} (here, \textit{i} represents $i^{th} MP$) the \textit{K}-most similar MPs using cosine similarity (steps 5 to 11). Then it map these similar MPs with their fWHR value in step 12. In step 13, the mean fWHR value among \textit{K}-most similar user is append to \textit{M}$_f$ for $i^{th}$ MP. The steps from 5 to 13 repeat for i = \{1, 2, ..., n\}, where ``n'' is the total number of MPs, and in step 14, the algorithm returns \textit{M}$_f$ as output. 

\begin{algorithm}
\SetAlgoLined
\textbf{Input: }MP-performance matrix V with n rows and m columns, \textbf{V}$\in R^{n\times m}$; MPs fWHR value array of size n, \textbf{M}$_p$; Number of nearest neighbors \textbf{k}, k$\geq$1.\\
\textbf{Output: }MPs fWHR value follow array of size n, \textbf{M}$_f$.\\
 \textbf{Initialization: }\textbf{M}$_f$=$[\ ]$\;
 \For{i = 1, 2, ..., n}{
  Initialization: \textbf{sim}=$[\ ]$\;
  $A$ = $V_{i}$, where $i$ is $i^{th}$ row\; 
  \For{j = 1, 2, ..., n}{
  $B$ = $V_{j}$, where $j$ is $j^{th}$ row\;
  append similarity($A$, $B$) in sim\;
  \begin{equation*}
similarity(A,B) = \frac{A.B}{||A||.||B||} 
\end{equation*}
\begin{equation*}
\hspace{3.5cm}= \frac{\sum_{l=1}^{m} A_l B_l}{\sqrt{\sum_{l=1}^{m} A_l^2}\sqrt{\sum_{l=1}^{m} B_l^2}}
\end{equation*}

  }
  Select the indices (except $i$) of k-most similar MPs from sim and store them in \textbf{idx}\;
  append MEAN($M_p[idx]$) in \textbf{M}$_f$\;
  
 }
 \caption{Feature Importance For Measuring Performance (FIMP)}
 \label{algo:HIM}
\end{algorithm}

During implementation, we utilizes the \textit{NearestNeighbors} function of \textit{scikit-learn} \cite{pedregosa2011scikit}. The main advantage of this functionality is that it is ideal for unsupervised nearest neighbors learning. The nearest neighbor method's principle is to find a predefined number of samples closest in terms of distance to the point under observation and predict the label from these. The number of samples (or \textit{K}) can differ depending on the local density of the points or specified by the user. For measuring the distance, we use cosine similarity as a distance metric.



\subsection{Finding K-value}\label{subsec:optimalK}
To find the optimal value for K, most traditional supervised machine learning task utilizes a dataset with target values. However, finding the optimal (or near to optimal) K value for the nearest neighbor algorithm is another challenge as in our case, the \textit{MP-performance} dataset lacks any ground truth. To identify optimal K-value, there are two popular (and standard) methods: (1) Selecting K as $\sqrt{n/2}$, where ``n'' is the total number of instances in the dataset \cite{nadkarni2016core, kodinariya2013review}. (2) Elbow method: This method's main principle is to increase K until it does not help to describe the data better \cite{kodinariya2013review, syakur2018integration}.

To overcome this challenge, it is essential to generate ground truth in the \textit{MP-performance} dataset. Since our analysis depends on MPs' voting activity, we have clustered these MPs based on their voting pattern using a community detection algorithm. Then, considering their community's label as ground truth, we find the optimal K-value using the elbow method. Next, we discuss the community detection algorithm we employed for identifying communities.

\noindent \textbf{Eigenvector Community Detection Method: }We use eigenvector community method (using \textit{igraph} function \textit{cluster\_leading\_eigen}) to identify communities based on voting pattern of MPs. The reason for selecting the eigenvector community algorithm is that eigenvector helps us answer: (a) Who or what holds wide-reaching influence; and (b) Who or what is important in the network on a macro level.

\begin{table*}
\centering
\begin{tabular}{|l|l|l|l|l|l|l|} 
\hline
                                         & Rada 4            & Rada 5            & Rada 6           & Rada 7           & Rada 8           & Rada 9            \\ 
\hline
Nodes                                    & 442          & 426          & 475         & 435         & 431         & 396          \\ 
\hline
Edges                                    & 96319        & 90478        & 109093      & 93865       & 91904       & 78529        \\ 
\hline
Average shortest path length             & 34.90723469  & 41.828876    & 20.3227093  & 6.409036496 & 10.04817353 & 32.30034522  \\ 
\hline
diameter                                 & 2            & 2            & 2           & 2           & 2           & 2            \\ 
\hline
Transitivity                             & 0.986060772  & 0.995219284  & 0.971641596 & 0.991958571 & 0.988744866 & 0.99911047   \\ 
\hline
Average CC                               & 0.272771409  & 0.351430933  & 0.225420981 & 0.26865168  & 0.141400213 & 0.224968922  \\ 
\hline
Edge dendity                             & 0.988282492  & 0.999480806  & 0.969069509 & 0.994385296 & 0.991787622 & 1            \\ 
\hline
Average degree                           & 435.8325792  & 424.7793427  & 459.3389474 & 431.5632184 & 426.4686775 & 396.6111111  \\ 
\hline
Total triangles                          & 40969557     & 37809336     & 48330207    & 39763563    & 38332623    & 30728196     \\ 
\hline
Number of connected components           & 1            & 1            & 1           & 1           & 1           & 1            \\ 
\hline
K                                        & 5            & 4            & 1           & 5           & 4           & 2            \\ 
\hline
t-value                                  & -0.662226601 & -1.879278323 & 0.53373401  & 3.313508591 & 1.377652 & 1.65   \\ 
\hline
p-value                                  & 0.507875362  & 0.060320276  & 0.593608934 & 0.000932114 & 0.1684 & 0.0988  \\ 
\hline
\end{tabular}
\caption{Network properties, K-value, and fWHR importance during RollCall for Rada 4 to 9.}
\label{tab:MPFollow}
\end{table*}

\begin{figure*}
\subfloat[Rada 4]{\includegraphics[width=0.3\linewidth]{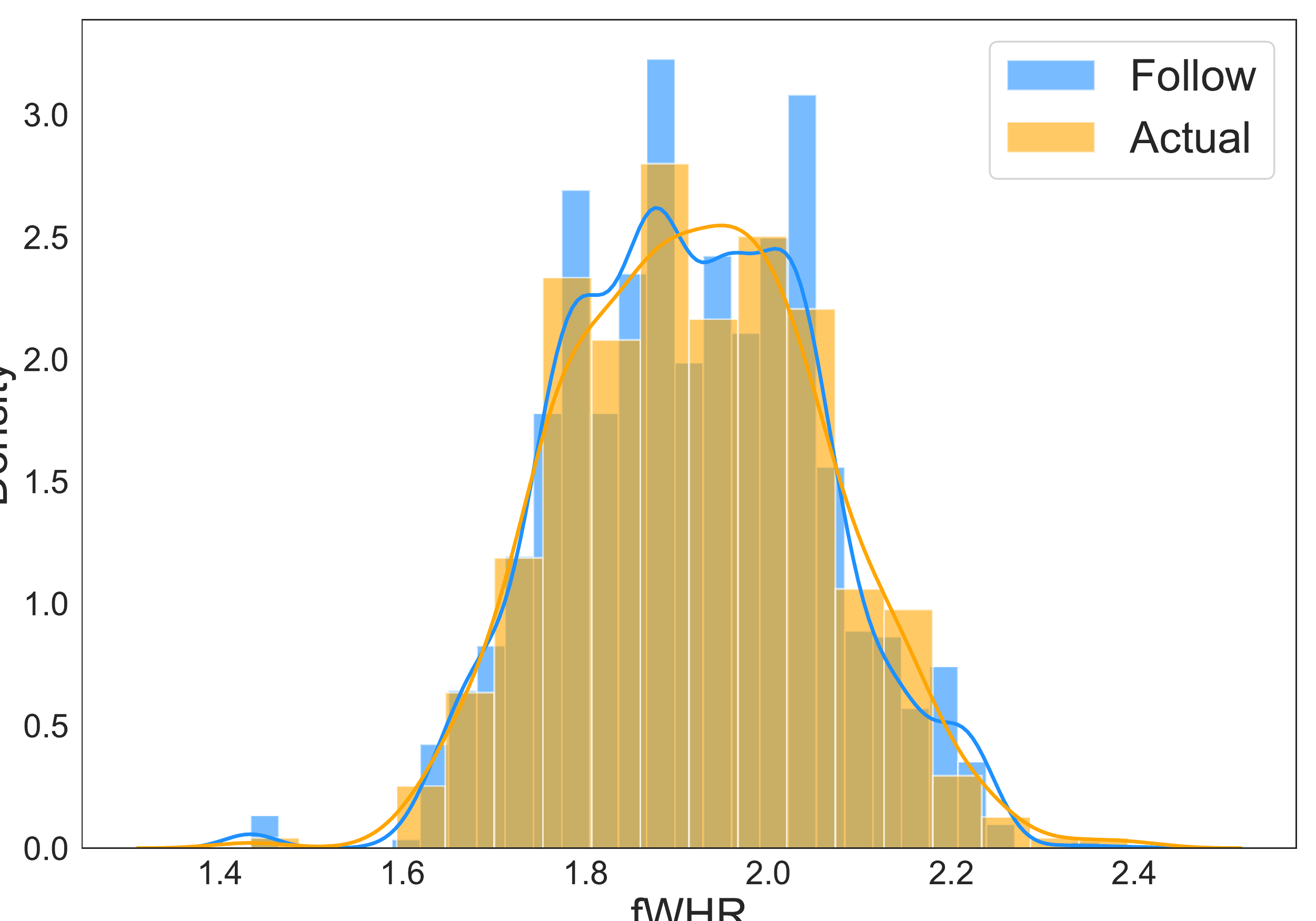}}
\hspace{3mm}
\subfloat[Rada 5]{\includegraphics[width=0.3\linewidth]{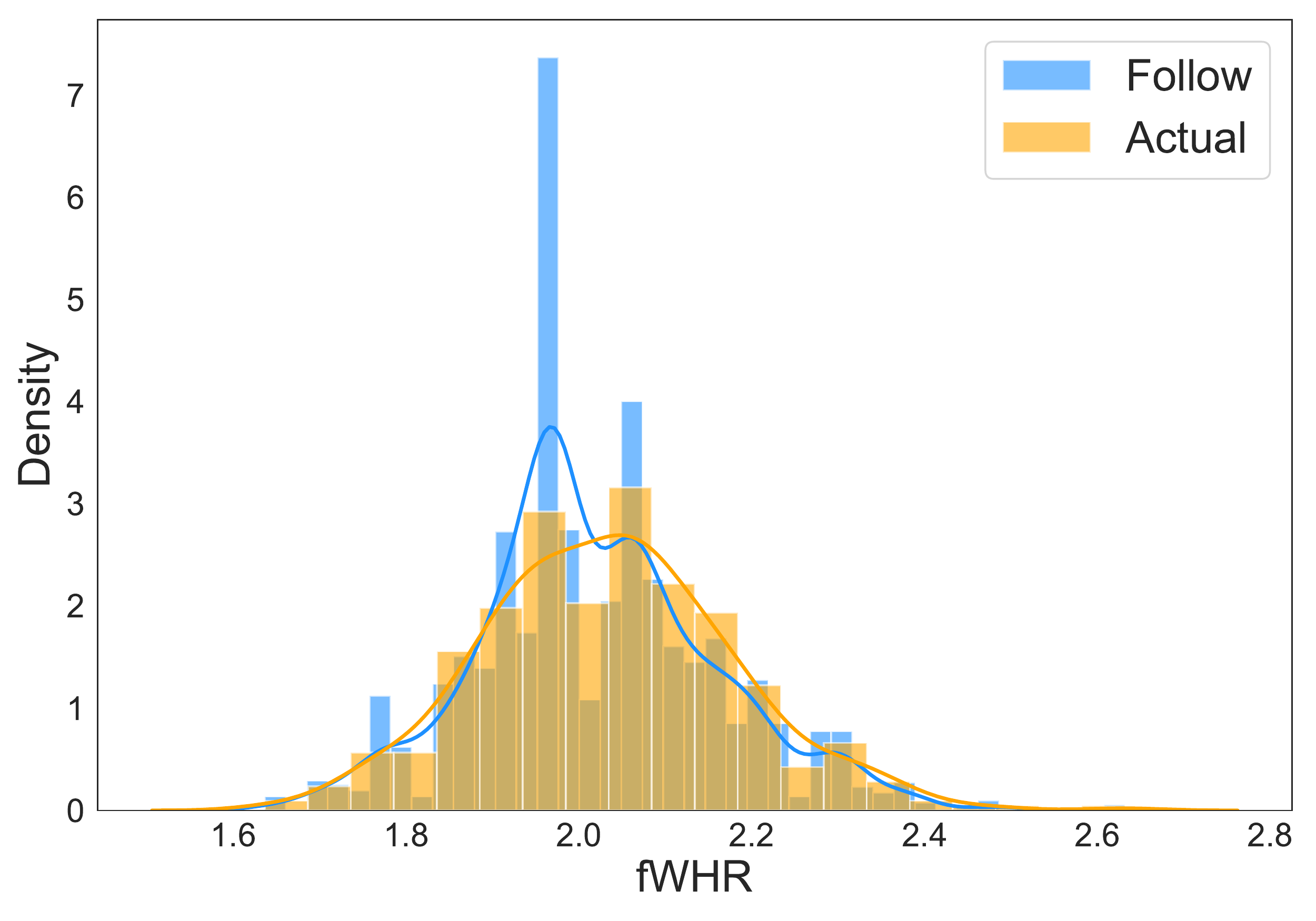}}
\hspace{3mm}
\subfloat[Rada 6]{\includegraphics[width=0.3\linewidth]{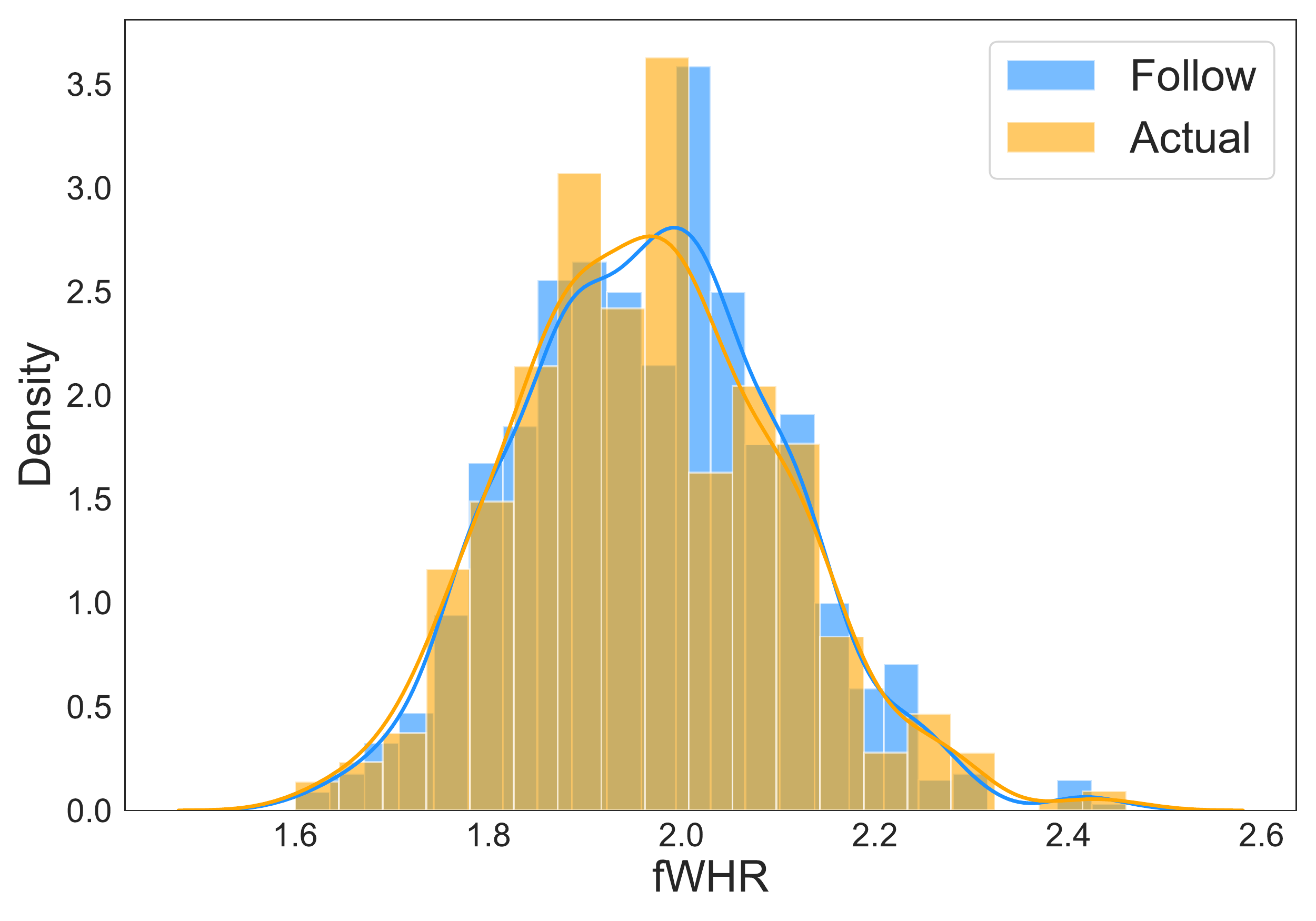}}\\
\subfloat[Rada 7]{\includegraphics[width=0.3\linewidth]{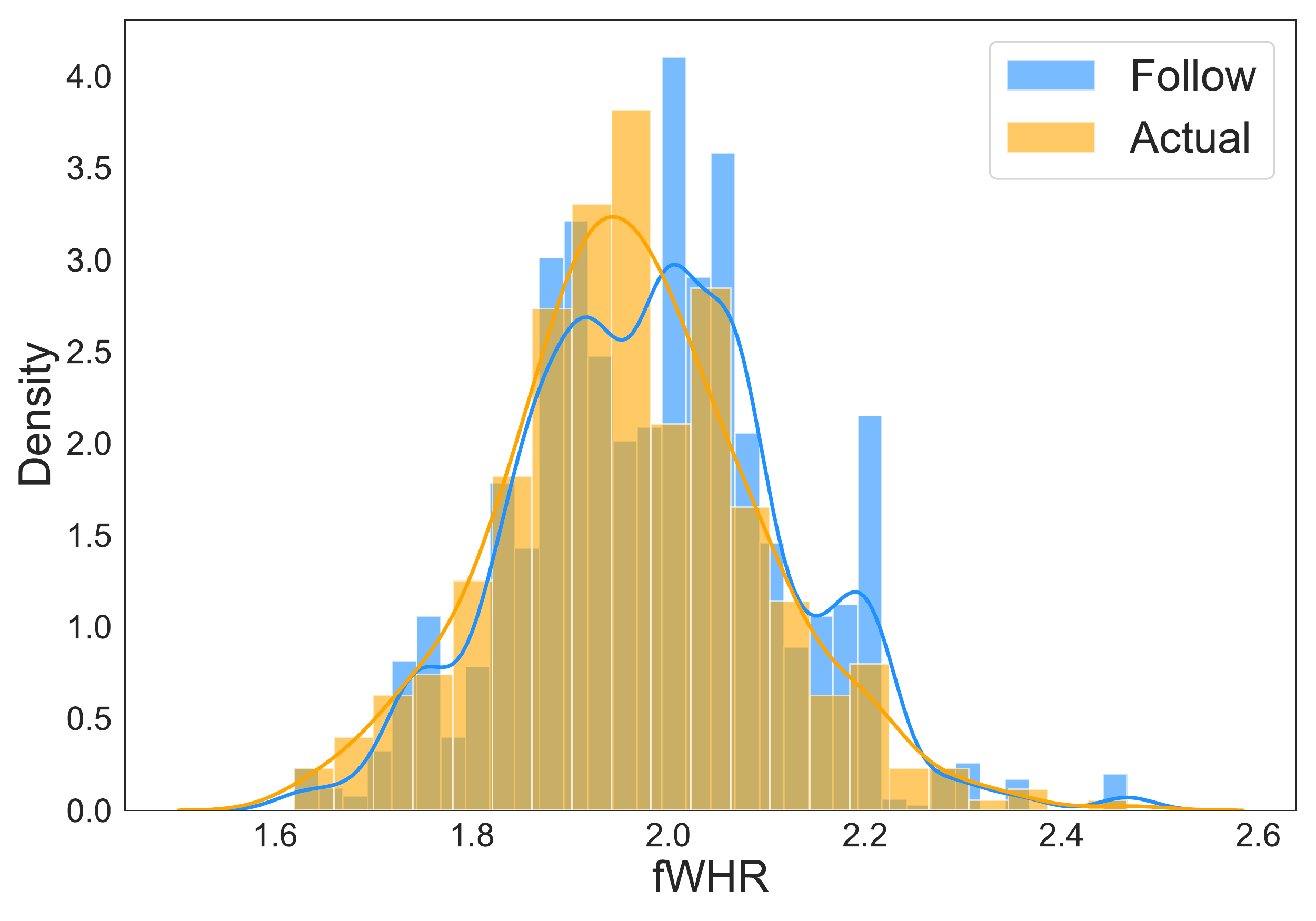}}
\hspace{3mm}
\subfloat[Rada 8]{\includegraphics[width=0.3\linewidth]{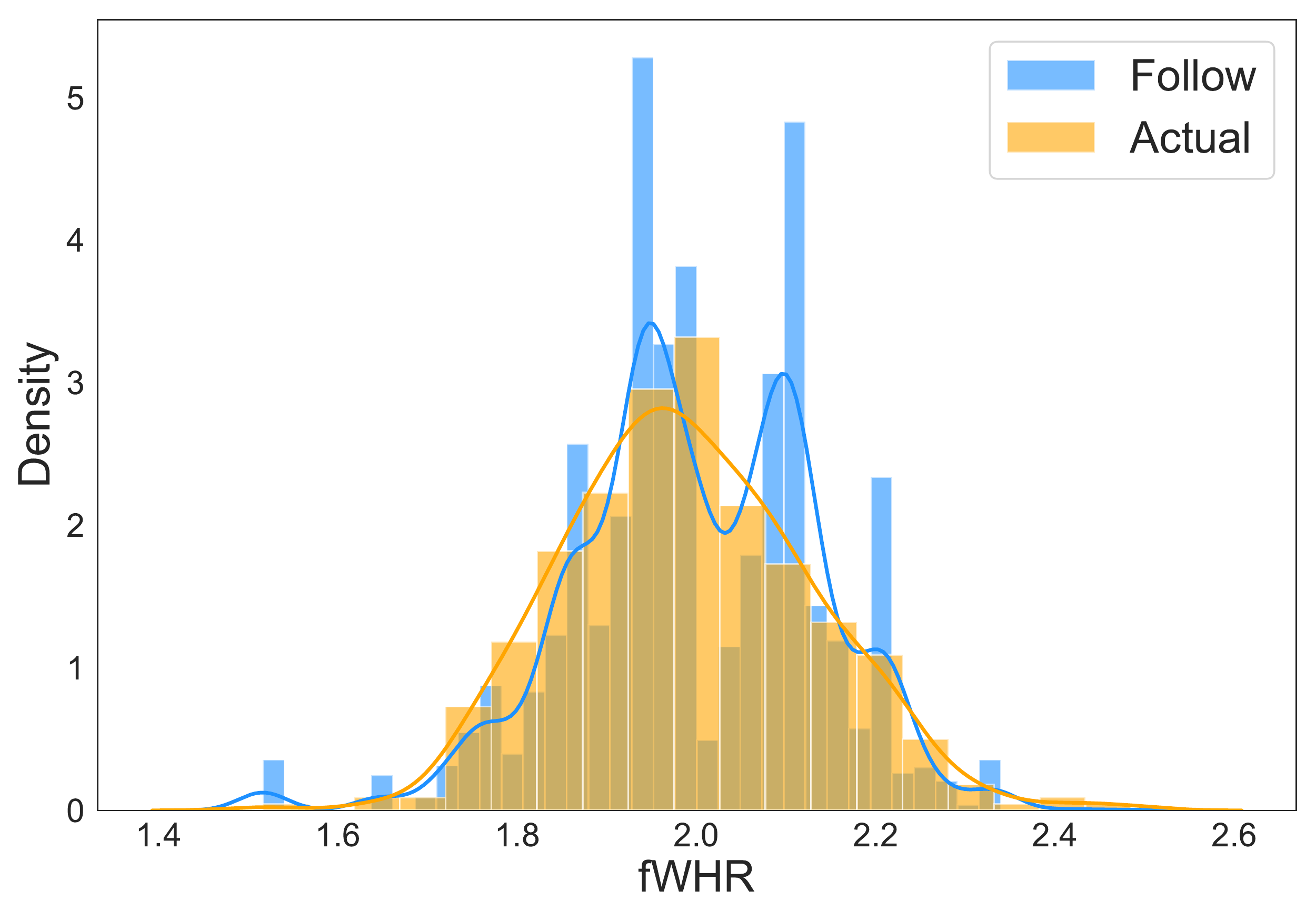}}
\hspace{3mm}
\subfloat[Rada 9]{\includegraphics[width=0.3\linewidth]{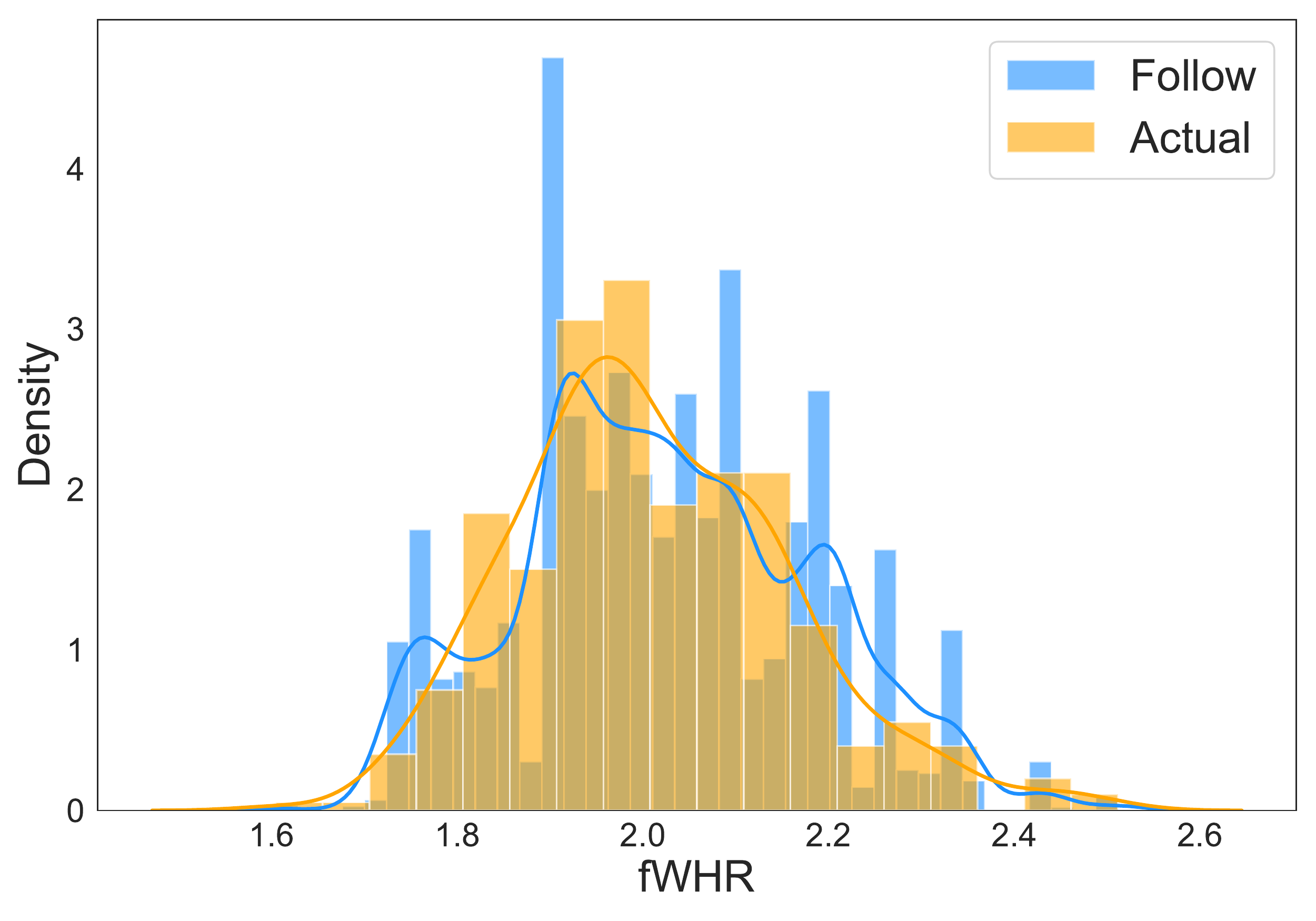}}
\caption{Density plot for MPs actual fWHR vs MPs follow fWHR during RollCall for Rada 4 to 9.}
\label{fig:MPfollow}
\end{figure*}

The eigenvector community method works by calculating the eigenvector of the modularity matrix for the largest positive eigenvalue and then dividing vertices into two communities based on the sign of the corresponding element in the eigenvector. If all eigenvector elements are of the same sign, that means that the network has no underlying community structure. Using this method, we find K's optimal value for all the Rada. 

\section{Results}\label{sec:Results}
We start our analysis by creating network between MPs based on their rollcall voting for each Rada. In the network, MPs represent nodes and their co-voting on bills represented as edges. Therefore, if MP \emph{A} and \emph{B} vote ``Yes'' on \emph{n} number of bills, then there exist a weighted edge between MP \emph{A} and \emph{B} with weight as \emph{n}.  Table \ref{tab:MPFollow} shows the network properties for Rada 4 to 9. Based on edge density, we can infer that all networks are approximately completely connected. This further supports the higher value of transitivity and presence of only one giant component across all networks.

Next, we use eigen vector community detection method to calculate number of community in each network. Considering the community label for each node as ground truth and RollCall votes as node features, we calculated optimal \emph{K} value for each network using elbow method (shown in Table \ref{tab:MPFollow}). 

Earlier in Section \ref{sec:datasetDescription}, we computed the fWHR value for each MP using their face images. Lets call this fWHR value $fWHR_{act}$ as it represents the actual fWHR value for MPs calculated using their faces. Now, as we already calculated \emph{K} value for each Rada, the \textit{FIMP} algorithm is used to identify the impact of fWHR in co-voting. As already mentioned in Section \ref{subsec:method}, the \textit{FIMP} algorithm output the mean fWHR value of \emph{K}-most similar MPs using cosine similarity for each MP. For convenience, we denote output of the \textit{FIMP} algorithm by $fWHR_{fol}$ as it is calculated using other MPs fWHR value that one follows. Next, we test whether $fWHR_{fol}$ and $fWHR_{act}$ comes from the same population. Considering that both $fWHR_{fol}$ and $fWHR_{act}$ are numerical variables, we perform t-test. We begin with the null hypothesis ($H_0$) and alternate hypothesis ($H_a$) as:


\begin{itemize}
    \item $H_0$: The $fWHR_{fol}$ and $fWHR_{act}$ comes from the same population.
    \item $H_a$: The $fWHR_{fol}$ and $fWHR_{act}$ comes from different population.
\end{itemize}

The t-test results p-value for each Rada (refer Table \ref{tab:MPFollow}). Considering $\alpha=0.05$, we observe that p-value is smaller than $\alpha$ only for Rada 7. Since, $\alpha$ $>$ p-value, we reject $H_0$ for Rada 7. Based on t-value and p-value, we can infer that at a 5\% level of significance, from the data, there is sufficient evidence to conclude that in rada 7, MP follows high fWHR MPs (see Figure \ref{fig:MPfollow}(e)). However, for other Rada (Rada 4, 5, 6, 8, and 9), we can not conclude the same. Figure \ref{fig:MPfollow} shows the actual and follow fWHR for all Rada. We can observe that actual and follow fWHR has approximately similar distribution for Rada 4, and 6. For Rada 5, 7, 8 and 9, distribution is somewhat different for actual and follow, which is also evident from low p-value for these Rada.

\section*{Limitations of the study}

We have to be careful when interpreting the results. Dominance, risk-taking, aggression, and achievement drive are not prerequisites for a political career. Yet, they can help when time is scarce, the level of uncertainty is high, and political competition induces information overload. In this case, leadership could play a significant role. What we cannot claim given the limitations of our data, is that MPs judge other MPs as leaders from their faces. In real life, people use a lot of information beyond our data, e.g., voice, facial expressions, jawlines, symbols, previous experiences, rumors, and reputation. We do not actually know if political leaders ``activate'' their leadership to influence others when they need it or they are perceived as more dominant and influential in the first place. In simple words, we do not know if leaders influence others or do others have a prior preference to follow leaders. Therefore, we acknowledge that our findings should not be interpreted in terms of causation. We believe that future research will benefit significantly from studying the dynamic of relationships between MPs: how exactly they ``activate'' their leadership clues and recognize leadership clues of others.

Previous research has shown that data-driven approaches to identify facial trait using images are not perfect \cite{stoker2016facial}. Some scholars use mouth width as an alternative metric of facial trait \cite{re2016big}. Moreover, other scholars suggest using machine learning algorithms to analyze the whole face \cite{turk1991eigenfaces, zhao2003face} instead of fWHR, which is the one-dimensional measure. Despite these caveats, our approach distinguished low and high fWHR politicians and predicted their voting behavior. 
Moreover, we performed a few steps of data quality check including face rotation in order to increase validity of our analysis. Since the distribution of fWHR in our sample is normal, we do not expect to have some systematic bias in our analysis (see Table \ref{table:DatasetAllRada}). 

\section{Conclusion}
\textbf{Summary of the results: }Drawing on the example of Ukrainian politicians and their co-voting from 2002 to 2019, we conclude that MPs tend to show no impact in their vote from colleagues with higher or lower fWHR. Considering the hypotheses, we proposed at the beginning of the text, \textbf{the data are not in line with the hypothesis of leader performance}. This hypothesis implies that politicians with higher fWHR has no impact whatsoever to able to summon votes from their colleagues because of their facial traits. 
However, we find evidence for the hypothesis of homophily, i.e., politicians prefer to follow colleagues with similar fWHR. 

\noindent \textbf{Contribution to the literature and new research directions: }This paper contributes to three streams in the literature. While previous studies of political leadership addressed electoral success and personal traits of politicians \cite{berggren2010looks,todorov2005inferences,lewis2012facial}, this paper looks at politicians' performance. Second,  while previous studies typically relied on respondents \cite{berggren2010looks,todorov2005inferences}, this paper explores a data-driven approach to measure the facial trait of politicians. Third, this paper connects the psychological literature of leadership and facial trait \cite{todorov2015social,zhao2003face} with the sociological literature of social ties and social networks \cite{jackson2010social, magelinski2018legislative, magelinski2019detecting, harder2019hybrid, harder2020predicting}. Finally, we suggest that future research of facial trait will benefit significantly from shedding more light on the social context of actual interactions between individuals and various sources of information they use to signal and recognize visual clues of leadership.


\section*{Acknowledgment}
This research is funded by ERDF via the IT Academy Research Programme and H2020 Project, SoBigData++, and CHIST-ERA project, SAI.

\bibliographystyle{IEEEtran}
\bibliography{Seeders}

\end{document}